# Grain refinement in unalloyed tantalum structure deposited using Wire + Arc Additive Manufacture and vertical cold rolling

G. Marinelli*[,a], F. Martina[a], S. Ganguly[a], S. Williams[a]

[a]Welding Engineering and Laser Processing Centre (WELPC), College Road, Cranfield University, Cranfield, MK43 0AL, UK

*Corresponding Author. E-mail address: g.marinelli@cranfield.ac.uk (G. Marinelli)

## Abstract

Components manufactured via Wire + Arc Additive Manufacturing are usually characterised by large columnar grains. This can be mitigated by introducing in-process cold-rolling; in fact, the associated local plastic deformation leads to a reduction of distortion and residual stresses, and to microstructural refinement. In this research, inter-pass rolling was applied with a load of 50 kN to a tantalum linear structure to assess its effectiveness in changing the grain structure from columnar to equiaxed, as well as in refining the grain size. An average grain size of 650 μm has been obtained after five cycles of inter-pass rolling and deposition. When the deformed layer was reheated during the subsequent deposition, recrystallisation occurred, leading to the growth of new strain-free finer equiaxed grains. The depth of the refined region has been characterised and correlated to the hardness profile developed after rolling. A reduction of porosity was also registered. Furthermore, a random texture was formed after rolling, which should result in isotropic mechanical properties. Wire + Arc Additive Manufacturing process demonstrated the ability to deposit sound refractory metal components and the possibility to improve the microstructure when coupled with cold inter-pass rolling.

## 1 Introduction

One of the most promising manufacturing technologies is Additive Manufacturing (AM) [1,2]. Three-dimensional structures can be promptly deposited starting from 3D-models, using a layer-by-layer approach [3]. The practical advantages are cost reduction, freedom of design and engineered mechanical properties [3]. Numerous researches have already discussed the potential of the AM of metallic materials, some of them are summarised in the work of Frazier [4].

Among the different AM techniques, wire-feed technologies, in particular the Wire + Arc Additive Manufacturing (WAAM) process, which employs an electric arc as the heat source, have already proven capable of producing large-scale components [5,6]. WAAM can directly fabricate fully-dense metallic large 3-D near-net-shape components with a much higher deposition rate, than most other metal additive manufacture processes [5,6], the highest rate so far being of 9.5 kg/h [7]. The WAAM process has successfully produced large-scale parts in stainless steel [8], Inconel ® [9], titanium [10], aluminium [11] and tungsten [12].

Furthermore, functionally graded structures of refractory metals have also been deposited using WAAM [13]. The manufacture of large and engineered components by WAAM is attractive also because of the low system and operating costs, as well as the modularity of the system design [6,14].

Tantalum is one of the most promising materials for high-temperature applications due to its high melting point and its inertness at high temperatures [15]. The stable oxide layer that immediately covers tantalum components is the main reason for its chemical stability against corrosion [16]. Furthermore, tantalum is characterised by a large ductility at room temperature, unlike other refractory metals such as tungsten and molybdenum [17,18]. These distinctive properties make tantalum a great candidate for the production of components for the aerospace sector as well as the defence, the electronics and chemical industries [15,17,19].

A few studies have been reported with regards to the development of an additive manufacturing process for unalloyed tantalum, mainly employing a laser as the heat source and tantalum powder as the feedstock. Laser-powder-bed-fusion (LPBF) has been used by Zhou et al. [20] to produce one of the first additively manufactured structures of unalloyed tantalum. In the work of Thijs et al. [21], the LPBF process was studied for unalloyed tantalum focusing mainly on the microstructural evolution and mechanical properties. Furthermore, tantalum coating has been successfully deposited on titanium using engineered net shaping (LENS) in order to produce controlled porous components with enhanced bio-properties [22]. A further study on porous tantalum parts is reported in the study of Wauthle et al. [23], in which LPBF was used effectively to produce porous tantalum implants with fully-interconnected open pores.

When processed by AM, most metallic materials present unique microstructural features such as large columnar grains that can lead to marked anisotropy [3,21,24]. Microstructural refinement in AM structures can be achieved in various ways. However, methods such as the optimisation of process parameters to change the solidification regime, or the alteration of alloy chemistry [25], are not suitable when the stability of the process must be controlled constantly, and the purity of the alloys used must be preserved. A viable option could be based on inducing plastic deformation within the deposited layer and subsequent deposition of a new layer (effectively equal to a local heat treatment) [24]. This has been already proven for Ti-6Al-4V, where the columnar prior-β-grains and the strong fibre texture have been completely eliminated [24].

For tantalum components, investigations regarding the effect of rolling, high-pressure torsion and equal-channel angular pressing on the microstructural refinement have already been reported [26–30]. In particular, ultrafine and nanostructured tantalum components can be produced by applying severe plastic deformation steps to a conventional coarse-grained structure, achieving nanometre-size grains [28,31]. Such fine microstructure is desirable; for instance, if tantalum is used as a sputtering target because a strong texture would lead to an inconsistent erosion rate through the surface of the target [32,33].

In this paper, the potential for microstructure refinement in tantalum by the addition of cold-work has been investigated. In particular, the aim of the study was to verify whether sufficient plastic strain could be introduced by inter-pass rolling and whether recrystallisation would occur during the subsequent deposition. The impact of the rolling load on the depth of refining and grain size has been evaluated and correlated with different hardness profiles measured in the structures. The influence of the recrystallisation on the developed crystallographic texture was also characterised.

## 2   Experimental Procedure

Unalloyed tantalum wire with a diameter of 1.2 mm was used for the Wire + Arc Additive Manufacturing process. A cold-rolled tantalum plate with a length of 210 mm, a width of 50 mm and a thickness 4 mm was used as substrate to start the deposition. The surface of the plate was ground and rinsed with acetone prior to the deposition to remove most of the contaminants. Materials' compositions are shown in **Table 1**.

**Table 1**: Elemental composition (wt.%) of the tantalum substrate and wire used in this study.

|  | W | Mo | Ta | Ti | V | Cr | Fe | C | N | O | K |
|---|---|---|---|---|---|---|---|---|---|---|---|
| Substrate | <0.05 | <0.05 | 99.99 | <0.05 | <0.05 | <0.05 | <0.05 | 33 ppm | <10 ppm | 60 ppm | <10 ppm |
| Wire | <0.05 | <0.05 | 99.98 | <0.05 | <0.05 | <0.05 | <0.05 | 36 ppm | 11 ppm | 190 ppm | <10 ppm |

**Fig. 1** shows the layout of the apparatus used for the deposition. The cartesian reference system used throughout the experiment is reported for all the schematics. A conventional tungsten inert gas (TIG) welding torch, a power supply and a controlled wire feeder were used for the deposition. The heat source, the wire delivery system and the substrate were attached to three linear motorized high-load stages assembled in XYZ configuration.

The direction of deposition was always kept constant for each successive layer and the wire was fed from the side of the weld pool. The apparatus was surrounded by a heavy-duty enclosure, purged with high-purity argon to keep the level of $O_2$ around 100 ppm. The linear structure was produced using the parameters shown in **Table 2**. A straight wall was produced with a length of 190 mm. All the samples analysed within this study were extracted from the same build.

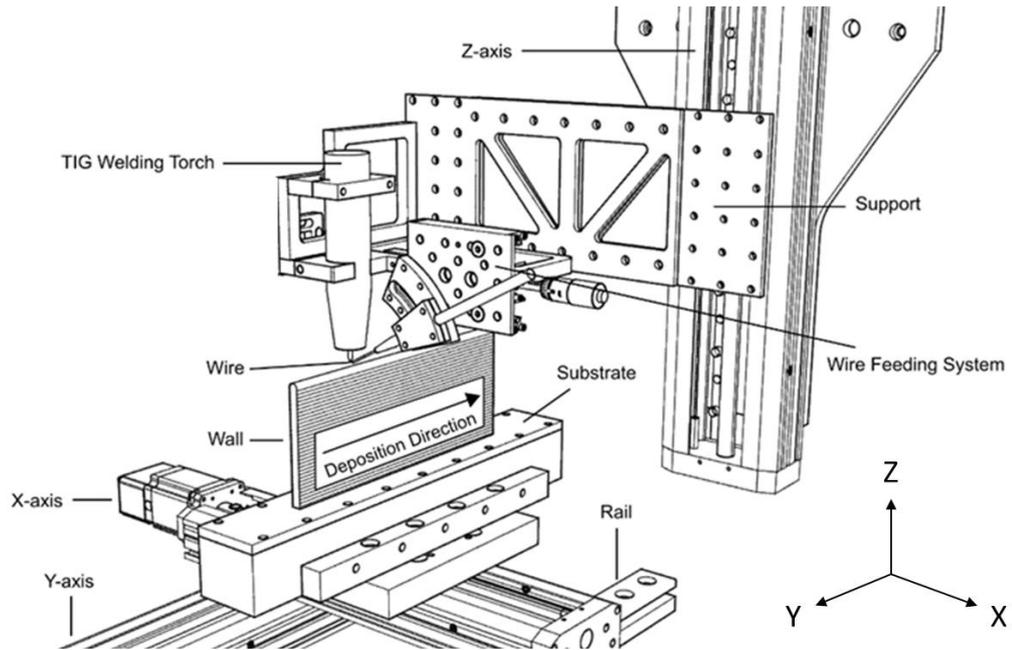

**Fig. 1**: Set-up used for the deposition of unalloyed tantalum using the WAAM process.

**Table 2**
WAAM process parameters used for the deposition of unalloyed tantalum.

| Parameter | Value |
|---|---|
| Travel Speed (TS) [mm/s] | 4 |
| Welding Current (I) [A] | 300 |
| Wire Feed Speed (WFS) [mm/s] | 40 |
| Shielding Gas Composition (SGC) [%] | 100 He |
| Gas Flow Rate (GFR) [L/min] | 15 |
| Oxygen Level [ppm] | ≈100 |

      A rolling rig was used for the application of inter-layer cold working. The roller travelled at a constant speed of 10 mm/s, applying a vertical load of 50 kN onto the structure deposited. A schematic diagram of the rolling equipment is shown in **Fig. 2**. The force applied by the roller was calibrated with a load cell before the experiments. Every rolling pass was carried out when the material had cooled down to room temperature.

      Please note that the specimen had to be taken out of the inert environment whilst still clamped to the backing bar, placed on the rolling rig for the rolling pass, and then returned to the enclosure for where the inert environment had to be re-created for the following layer deposition, which started only after the oxygen content was below 100 ppm. Rolling and deposition had to be done on two different set-ups because creating an inert atmosphere around the entire rolling rig was not practical.

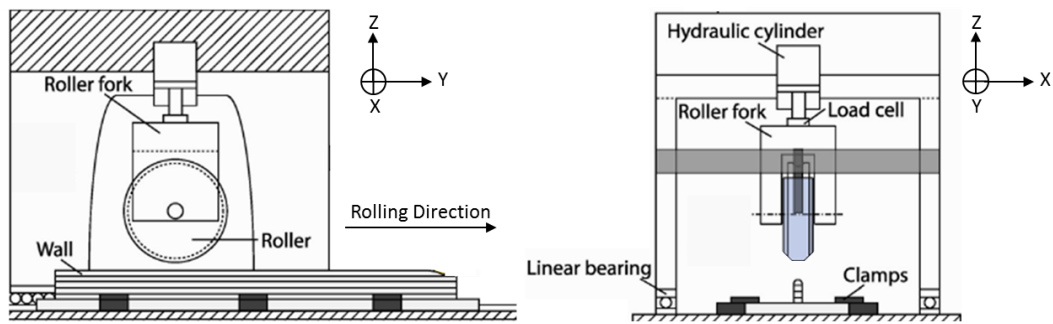

**Fig. 2**: Schematic of the rolling rig used for the application of cold working between each successive layer.

The location and the manufacturing route for each sample analysed are shown in **Fig. 3.** Four main samples were produced using a different combination of deposition and rolling. The sample called 1R was manufactured only after the deposition of 15 layers of unalloyed tantalum without the application of any rolling steps and served as control specimen. The sample 2R was extracted from the portion of the wall that was normally deposited for 15 layers and just rolled once, at the top. The sample 3R had the same history of 2R, except a 16th layer was deposited on top of the rolled one. The sample 4R had 15 layers of the deposition without any rolling, after which 5 more layers were deposited with inter-pass rolling applied. The regions from which sample 1R, sample 2R and sample 3R were extracted were 40 mm long and the sample 4R was extracted from the remaining 70-mm-long structure. The structure was produced gradually decreasing the deposition and the rolling path length.

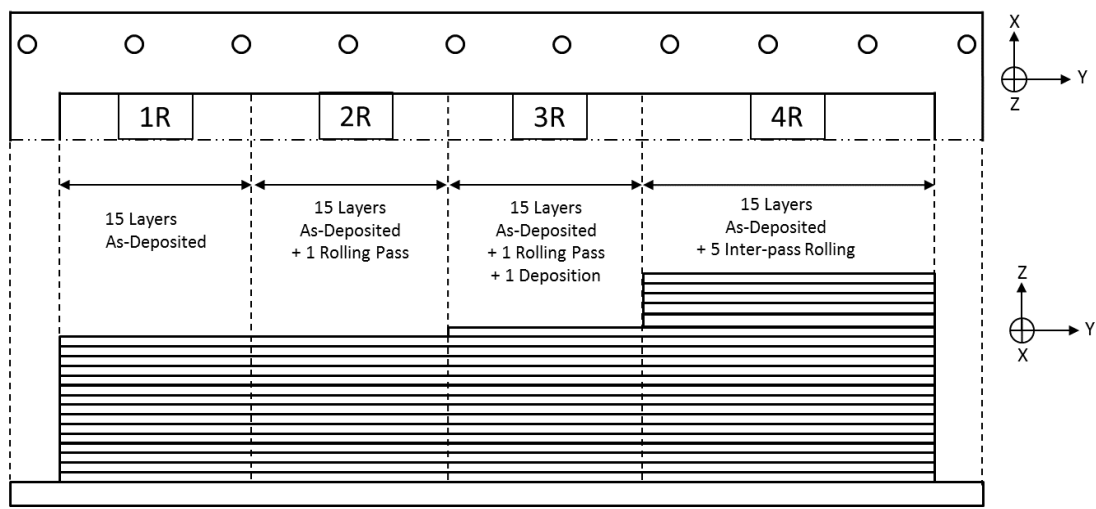

**Fig. 3**: Schematic of the manufacturing route used to produce each sample analysed. The name of the samples extracted per each region is also reported.

The samples were ground, polished and etched, and analysed for porosity, hardness, grain size and texture. The microstructure was examined using the cross-section perpendicular to the deposition direction. Vickers hardness contour maps were produced using the measurements collected using an automatic hardness testing machine. The main parameters for the hardness acquisition were 500g load and 10 seconds dwell time for each testing point. Several line-scans with 1 mm spacing were performed throughout the height and width of each cross-section to obtain the contour maps of hardness. The microstructure of the sample 4R was characterised by scanning electron microscope (SEM) operating at 20 kV. Grain size and grain orientation were measured with the aid of an electron backscatter diffraction (EBSD) detector within the SEM. Polar contour maps were used to evaluate the texture. The sample 4R was chosen as it was the sample that showed the most evident contrast in microstructure within the same build. The root and the top of the samples were measured to understand the variations within the microstructure.

## 3 Results

*3.1 Microstructure*

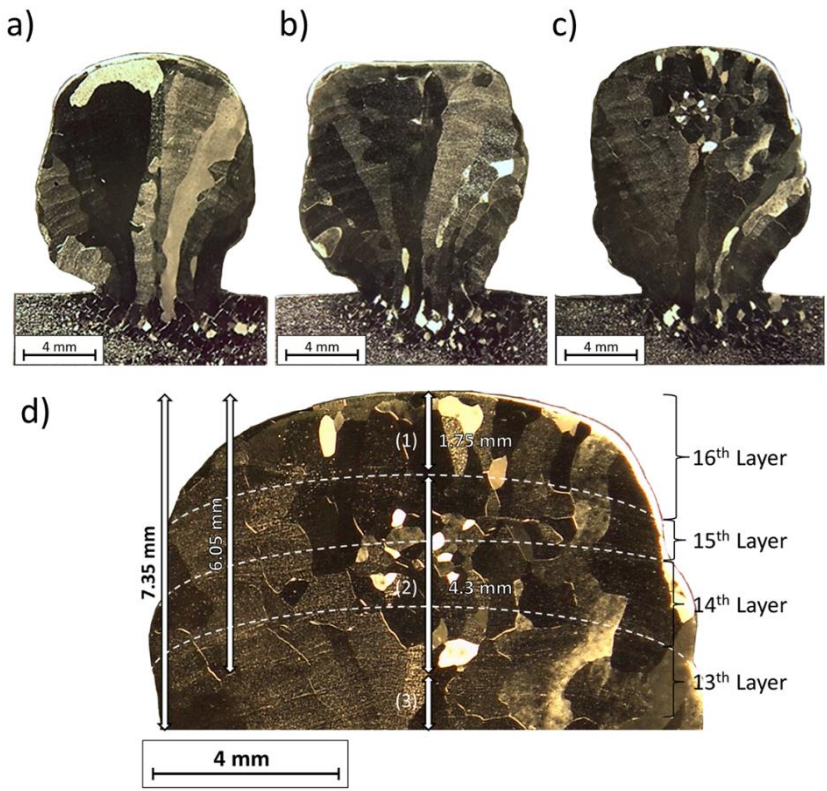

**Fig. 4**: Optical micrographs of sample 1R (a), 2R (b) and 3R (c). High magnification of the recrystallised top part of sample 3R (d).

The microstructure of 1R, 2R and 3R are reported in **Fig. 4a**, **Fig. 4b** and **Fig. 4c** respectively. All the samples produced presented a smaller width near the baseplate; this has been explained elsewhere [34]. The sample 1R (**Fig. 4a**) demonstrated large columnar grains, which grew epitaxially from the base plate toward the top of the sample parallel to the building direction, with dendrites that were 1-mm-wide on average and extended from the base plate to the top of the deposit. For 2R, the size and the morphology of the grains were essentially the same as 1R as the only rolling pass just flattened the upper surface, plastically deforming the structure (**Fig. 4b**). The sample 3R (**Fig. 4c**) presented a significant reduction in size and a change in morphology for the grains localised in the upper region. The majority of the refined equiaxed grains had a size between 250 μm and 350 μm. There were a few larger grains which had a size between 600 μm and 850 μm.

**Fig. 4d** shows a higher magnification micrograph of the refined region of sample 3R. The lines identifying the individual layers could be traced thanks to the geometrical features of the outer surfaces. The last four layers were chosen to illustrate the depth and width of the refined region. In particular, the top 7.35 mm of the structure were reported, and three different zones of the microstructure can be distinguished: the last deposited layer (1) which is effectively in the as-deposited conditions, with fine columnar grains growing epitaxially from the refined grains below; the refined region (2) which extended from 1.75 mm in depth until 6.05 mm in depth, across three layers below the last one (from the fifteenth to the thirteenth layer) and was localised at the middle of the structure; and the layers (3) where the effect of rolling was not visible and where the large columnar grains were left unchanged.

The microstructural features of the sample 4R are reported in **Fig. 5**. In particular, **Fig. 5a** shows the entire cross-section, with a marked change in microstructure for the upper layers. The boundaries of each layer were not clear due to the considerable lateral and vertical deformation given by inter-pass rolling. This also resulted in an average larger width for the upper layers. The large columnar grains typical of the as-deposited condition extended from the substrate until 7.25 mm in height.

The refined region extended for around 6.4 mm in height and it had a total width of around 7.6 mm. Average grain size was around 650 μm. The top 2.2 mm presented a typical microstructure of the as-deposited layer without any rolling after deposition, similar to the top of sample 3R. **Fig. 5b** and **Fig. 5c** report details of the microstructure of the refined and columnar regions with highlighted grain boundaries. It is important to note the change in scale.

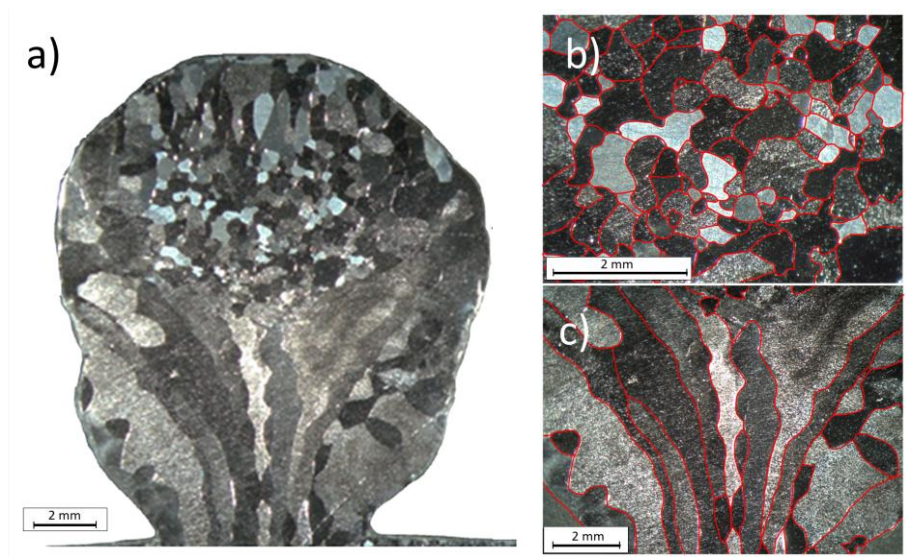

**Fig. 5**: Optical micrograph of sample 4R (a). Detail of the microstructure from the recrystallised zone (b) and from the columnar as-deposited region (c) with highlighted grain boundaries.

*3.2    Porosity and deformation bands*

**Fig. 6** reports the optical images of the polished surface for samples 1R, 2R and 3R. **Fig. 6a** shows some micron-size pores in sample 1R. A similar distribution of porosity was found in sample 2R near the top and bottom of the structure (**Fig. 6b** and **Fig. 6c** respectively). **Fig. 6e** shows the porosity found in the bottom part of sample 3R and it resulted to be similar to sample 1R and sample 2R. **Fig. 6d** reports, in particular, the boundary between the sixteenth layer as deposited and the fifteenth layer of sample 3R, which has been deposited, rolled and refined. It is clear that the refined region had smaller and fewer pores that the last as-deposited layer, which was characterised by the presence of larger porosity like samples 1R and 2R.

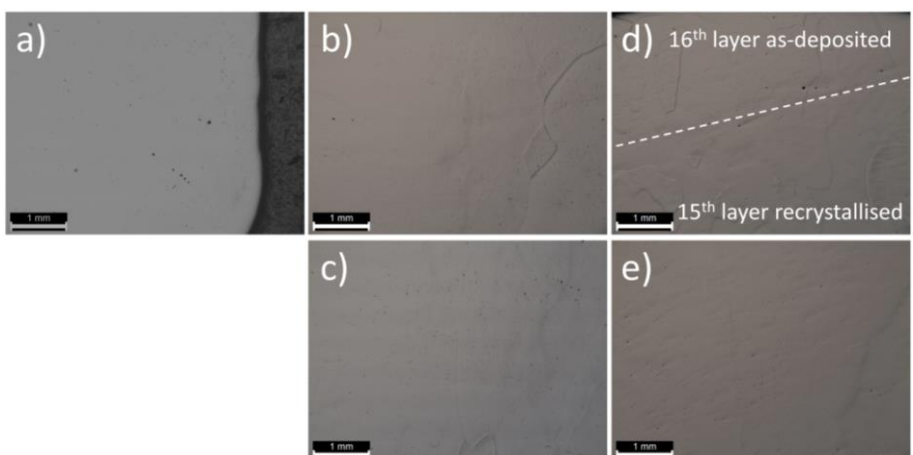

**Fig. 6**: Polished cross-sections showing the porosity within the sample 1R (a), the top of the sample 2R (b), the bottom of the sample 2R (c), the top of the sample 3R (d) and the bottom of the sample 3R (e).

**Fig. 7a** report some elongated pores found on the top of sample 2R. **Fig. 7b-d** show some distinctive features found on the surface of sample 2R within the same grain and across multiple grains. In particular, in **Fig. 7c** and **Fig. 7d** is shown an unusual banding seen across some grain boundaries. Each line of the bands was pointing in the direction of around 30° to 45° compared to the Z-axis. These deformation bands were associated with a high degree of lateral deformation.

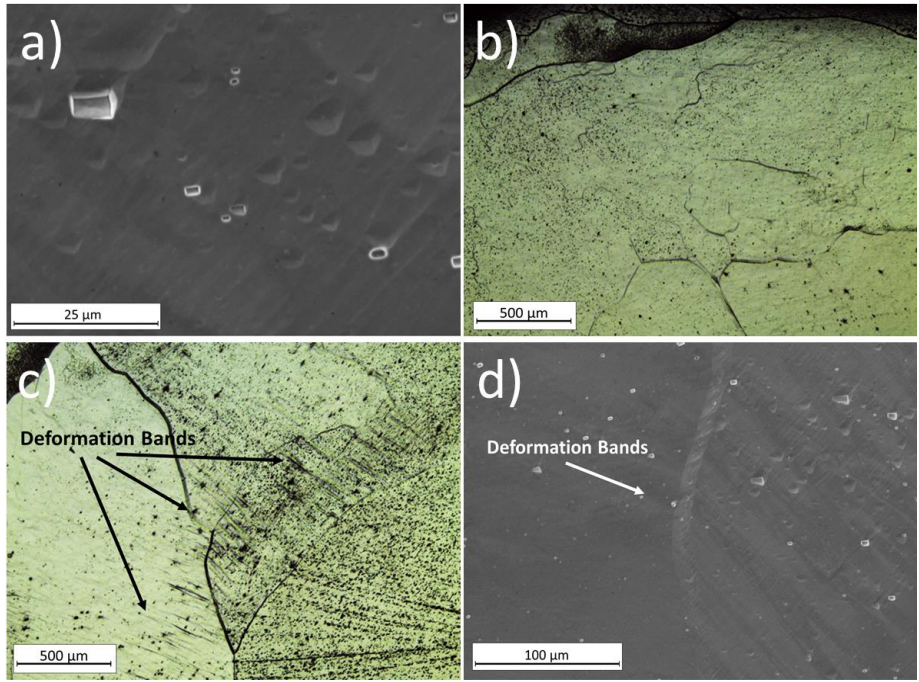

**Fig. 7**: Flattened pores found at the top of the sample 2R (a); Microscopic bands at the top of the sample (b) and across grains in sample 2R (c-d).

*3.3   Hardness*

**Fig. 8a** reports the hardness distribution for the sample 1R, which represents the as-deposited material (control). The hardness resulted to be consistent, with small fluctuations around a mean value of 114 HV across the width and height. **Fig. 8b** reports the hardness map for the sample 2R, after the single rolling pass. The impact of rolling on the hardness values and distribution can be seen clearly. The application of rolling increased the hardness of the whole structure down to 10 mm from the top. A predominant high-hardness region developed from 1.0 mm to 6.5 mm in depth, with a width of around 4.0mm. The hardness almost doubled reaching the value of 205 HV at its peak. **Fig. 8c** shows the hardness distribution of the sample 3R. The overall hardness of the structure was reduced compared to the rolled structure. From the top, down to a depth of 1.75 mm, the hardness reduced and was similar to the average hardness of 1R, the as-deposited material. This means that the deposition of a new layer effectively annealed the underlying material.

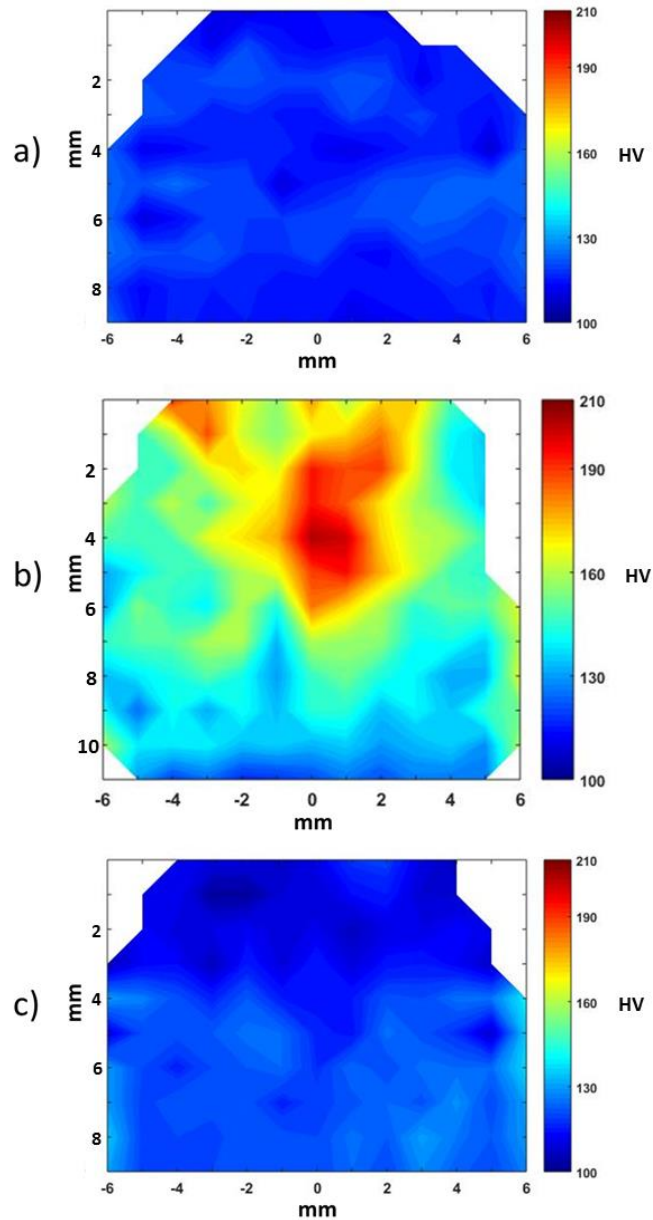

**Fig. 8**: Contour maps relative to the Vickers hardness measurements for sample 1R (a), 2R (b) and 3R (c).

*3.4   Texture*

The images reported in **Fig. 9** are maps acquired at the columnar and refined regions of sample 4R. The orientation of the crystals is indicated in the direction parallel to the building direction (Z-axis). The columnar region (**Fig. 9a**) was characterised by the absence of columnar grains oriented in the <101> direction, showing some preferential directions of crystallisation. The EBSD map for the refined region (**Fig. 9b**) presents much more refined grains with random orientation. Here, the columnar grains were completely refined to equiaxed grains, and the preferred orientation of crystals developed during deposition was eliminated.

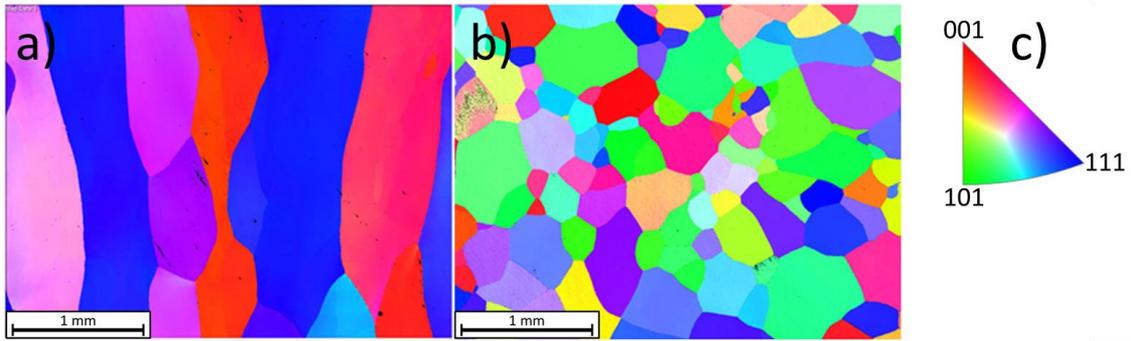

**Fig. 9**: EBSD images of the columnar as-deposited region (a) and the recrystallised grains (b) of sample 4R. Legend with regards to colour and orientation (c).

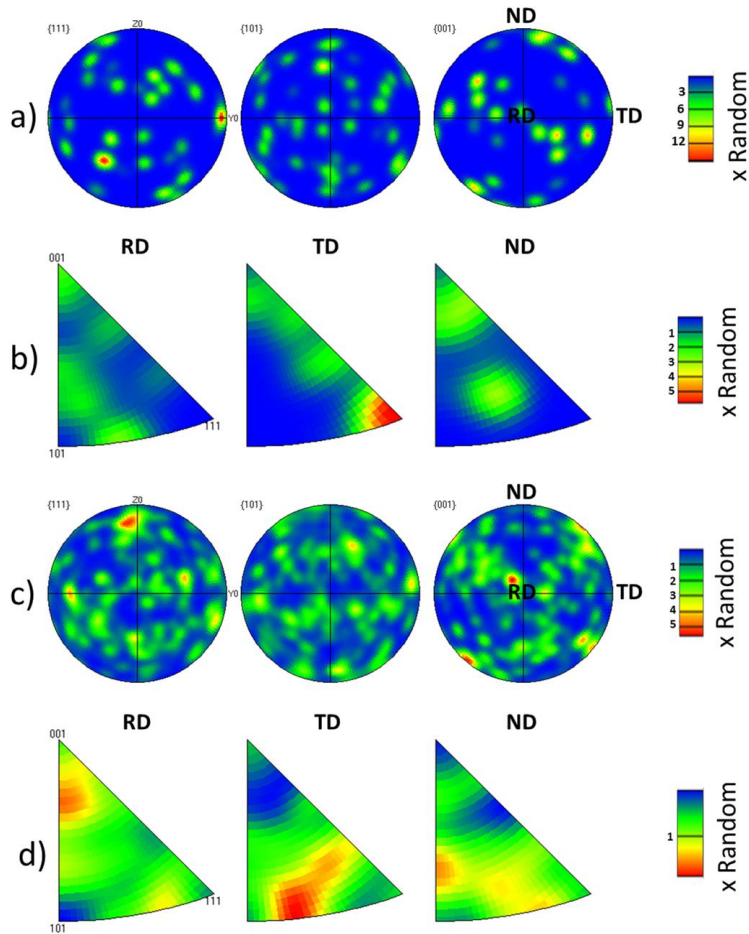

**Fig. 10**: Pole figures and inverse pole figures obtained from EBSD maps for the columnar as-deposited region (a-b) and the recrystallised grains (c-d) of sample 4R.

When comparing **Fig. 10a** and **Fig. 10c**, the change in texture due to rolling is evident. The columnar as-deposited region showed pole intensities peaking at 12 x random, while for finer equiaxed grains the peak intensities were around 5 x random. A marked scatter of the signal from different crystallographic planes was also clear for the refined region. It is important to note that the columnar grains were not characterised by a strong texture but showed a preferred orientation in the <111> direction (**Fig. 10b**).

# 4 Discussion

In all industrial sectors where tantalum finds application, a fine equiaxed microstructure with the reduction or elimination of the crystallographic texture is desirable, so as to benefit from improved mechanical properties and complete isotropy. The results of this study surely prove that in-process rolling leads to a large refinement of the grain structure, as well as reduction of porosity and texture. In the samples studied, the rolling load applied caused a compressive strain within the previously-deposited layers and the thermal field of the subsequent deposition allowed new recrystallised grains to form and grow. The recrystallisation occurred within the centre of the deposited structure, where the peak of strain was located and was extended as deep as three layers below the last deposited one. Successive steps of rolling and deposition led to a progressively larger area of fine equiaxed grains with an average grain size of 650 μm after five inter-pass rolling steps.

*4.1 Induced strain and the depth of recrystallisation*

4.1.1 Recrystallisation of unalloyed tantalum

The process that leads to the formation of new strain-free grains, which nucleate from deformed parent grains, is defined as recrystallisation. The formation of these new grains is driven mainly by the energy stored within the metal from the deformation; and by the temperature, as such process is a thermally-activated mechanism. In general, the deformation increases the density of crystal imperfections, as for instance dislocations, which act as nucleation sites for new grains when the recrystallisation temperature is reached. Thus, temperature and strain regulate the rates of nucleation and grain growth.

Unalloyed polycrystalline tantalum is characterised by a recrystallisation temperature that varies within the range from 900°C to 1450°C, depending on the level of impurities and, predominantly, on the level of strain/deformation [26]. Usually, a high content of impurities increases the recrystallisation temperature. Köck et al. [15] reported that the deformation of arc-melted tantalum components can be pronouncedly influenced by metalloids such as carbon, nitrogen and oxygen. In the work of Mathaudhu et al. [35], the reported recrystallization curves showed that all tantalum specimens were fully-recrystallized at 1100°C, but some of the samples recrystallised between 900°C and 1000°C, depending on the degree of deformation applied. A typical range of strain level in unalloyed tantalum, in order to obtain recrystallisation, was indicated to range from 1.2 to 4.6 [36]. The recrystallisation temperature associated to different strain levels was also indicated: samples with a strain of 1.2 recrystallised at 1100°C; those with a strain of 2.32 recrystallised at 1000°C; and those with a strain of 3.48 recrystallised at 900°C [36].

4.1.2 The relation between hardness and strain

As the application of strain gauges on the external surfaces of the wall was not possible due to the surface roughness and the extremely high thermal field given by the deposition, the hardness profile was used to evaluate the strain distribution within the bulk of the structure. This because it is well known that tantalum components are subjected to considerable strain hardening, when cold working is applied, due to a high dislocations density produced by the plastic deformation [36,37]. Thus, the change in hardness after rolling can be directly associated with the strain.

In the work of Aditya et al. [27] on cold rolling of tantalum sheets, a trend between cold reduction (strain) and Vickers hardness was clearly shown. An increment in average hardness was seen until cold reduction superior to 65%, where the hardness stabilised around 200 HV. In the work of Mathaudhu et al. [35], the hardness of an arc-melted tantalum ingot was reported to increase after rolling. In particular, a hardness of 170 HV was reached after one rolling cycle and a hardness of 210 HV after two rolling cycles. In other studies on rolling of tantalum plates, micro-hardness values reached maximum levels around 230 HV after only two rolling cycles with a strain of 2.3 [30]. In general, a plateau of hardness was always seen around 210 HV - 220 HV for the studies available in the literature, even for a high level of deformation. This saturation in formability and hardness is associated with the high density of dislocations, which obstacle their own further motion [27].

4.1.3 The relation between hardness and dislocation density

An empirical relation has been proposed in the literature [38] which correlated dislocation density (ρ) to the variation of Vickers hardness for cold worked tantalum. The equation is reported as following (**Eq. 1**):

$$\rho = 1.6 \times 10^9 (m^{-2}) \left(\frac{H - H_{ing}}{MPa}\right)^2 \quad\quad\quad\quad (1)$$

Where H is the Vickers hardness after cold working and $H_{ing}$ is the hardness of the as-received material prior to deformation. Stuwe et al. [38] also proposed a minimum critical dislocation density in order to have spontaneous nucleation, which was indicated to be equal to $3 \times 10^{14}$ m$^{-2}$. From the equation reported, the corresponding Vickers hardness to the critical dislocation density is equal to 130 Hv [27,38]. When considering the hardness distributions reported in **Fig. 8**, it is possible to assess that the rolling step clearly induced a certain amount of strain within the bulk material associated with an increase in hardness to values as high as 210 HV. The deformation localised around the hardness maximum should also be considered characterised by a considerable dislocation density, enough to induce recrystallisation. Furthermore, the hardness contour map reported in **Fig. 8b** for samples 2R can be clearly considered a representation of the prior history of plastic deformation within the height and the width of the sample.

4.1.4 Effect of the thermal field on the hardness

The measurements of the thermal field evolution during and after the deposition have not been recorded with the use of thermocouples. The difficulties faced during the preparation of the experiments were the selection of thermocouples able to record a range of temperatures as high as 3000°C and also capable of being attached to the tantalum surface. The oxide layer on the tantalum surface made the attachment of the thermocouples difficult. Temperature measurements and thermal 3D-modelling would be needed for further developments of the study. This would help to understand if the depth of recrystallisation was mainly influenced by the strain or by the temperature. Although the lack of thermal data, the grain refinement achieved in sample 3R shows that the thermal field developed after the deposition clearly covered the range of recrystallisation temperature.

The contour map of sample 3R (**Fig. 8c**) shows how the thermal field developed after the deposition affected the hardness of the entire structure. There were particularly three main regions characterised by the hardness and the location. These regions corresponded precisely with the three regions of microstructure reported in sample 3R (**Fig. 4d**).

The region at the top extended from the top until 1.75 mm in depth which corresponded precisely to the fusion boundary location of the last layer (**Fig. 4d**). The hardness was found to be very close to the as-deposited material.

The region below this one corresponded entirely to the recrystallised region. The hardness was correlated to the size and shape of the new recrystallised grains. Even if the strain was almost totally relieved by the formation of these new defect-free grains, the density of grain boundaries locally increased when compared to the parent columnar grains. The grain boundaries impeded the dislocation movements. This has led to a local strengthening effect according to the Hall-Petch relation.

Additionally, the lower part of the sample 3R was characterised by a general decreasing in hardness when compared with the same portion of sample 2R but showing visible retention of strain. The change in strain concentration and the dislocation density induced by the rolling was not sufficiently high to induce recrystallisation so that recovery occurred. The small driving force, in terms of strain and temperature, led to a general softening without the formation of new grains.

4.1.5 Dislocations density and oxygen content

The critical hardness value of 130 HV obtained by the relation indicated by Stuwe et al. (**Eq. 1**) has been achieved for the top 8.0 mm of the sample 2R. Despite the dislocation density of $3x10^{14}$ $m^{-2}$ associated with this hardness value and the assertion that this is the minimum value to obtain nucleation of new grains, the recrystallization occurred only within a region characterised by a hardness ranging between 170 HV to 205 HV. The dislocation densities associated with these values of hardness should result to be respectively $5x10^{14} m^{-2}$ and $1.2x10^{15} m^{-2}$ using the relation of Stuwe et al. (**Eq. 1**).

This equation does not take into account the effect of interstitials on the hardness in tantalum. In fact, it is known that nitrogen and oxygen are strong hardeners of unalloyed polycrystalline tantalum [39]. In the work of Stecura [40], it is shown that the hardness of tantalum increased almost linearly when increasing oxygen content in solid solution. Samples with a different concentration of oxygen have been tested finding a hardness value of around 120 HV for a tantalum specimen with a purity of 99.95 wt.% and a hardness value of around 500 HV at about 1000 ppm of oxygen. The hardness of 114 HV for the as-deposited differs from the typical hardness of a tantalum structure of 90 HV mainly due to the 226 ppm of oxygen within the wire (**Table 1**).

The small amount of oxygen within the material used in this study could have led to a more pronounced strain hardening so that a higher nominal value of hardness was measured after cold working. Thus, the dislocation density for each region of sample 2R should be considered slightly lower than the over-estimation from the **Eq. 1**.

*4.2 Recrystallisation mechanism and Texture*

When using Wire + Arc Additive Manufacture to deposit unalloyed tantalum, columnar grains develop during the solidification and grow in the direction perpendicular to the deposition direction. This is a typical example of epitaxial growth along the building direction that some materials develop when processed with an additive manufacturing process as seen by Thijs et al. for unalloyed tantalum [21] and by Martina et al. for Ti-6Al-4V [34].

In the current study, the rolling load was applied in the direction parallel to the major axis of the columnar grains of the tantalum structure. Under the compressive stress, the columnar grains were deformed not only in the vertical direction but also laterally. This led to the formation of deformation bands among multiple grains with an angle close to 45° compared to the direction of the load applied (**Fig. 7c-d**).

Similar features have been seen from Mathaudhu et al. [30] and they have been addressed to the occurrence of slip. Nemat-Nasser et al. [41] affirmed that the slip of the screw dislocations on {110} planes was the dominant deformation mechanism observed in tantalum. Furthermore, elongated dislocations cells with small misorientations were found to form bands which were defined as the predominant microstructural characteristic of the shear-localization region. Micro-bands have also been reported in tantalum and their existence was mainly associated with the primary slip system in a <111> crystal direction [42]. For bcc metals, such as tantalum, commonly slip occurs in the <111> direction on the planes {110}, {112} and {123}, which are the planes with largest interplanar spacing [28]. The Burgers vector is a/2<111> [43] and these particular screw dislocations control the most of the deformation properties of bcc metals so that in tantalum. In the study of Liu et al. [42], it has been seen that grains in <111> direction stored a higher amount of energy compared to the other directions and also <111> grains were the preferential place for nucleation of recrystallised grains after application of thermal energy. They have also seen that the dislocations were distributed uniformly only in <111> grains with a deformed matrix and concluded that this evenly stored energy represented the strong driving force for the growth of new crystals [42].

Although this value should be treated with caution as the mapped area was relatively small and the grain size was large, the as-deposited columnar grains shown in this work presented some preferred orientation in the direction <111> (**Fig. 9a**, **Fig. 10a-c**).

The deformation applied by the roller led to the development of deformation bands, potentially series of slips, which have acted as nucleation sites for recrystallisation. The investigation needs further analysis in order to understand which family of crystal imperfections actually acted as a nucleation site for the new strain-free grains. Furthermore, the randomisation of the texture after recrystallisation was clear when analysing **Fig. 9b**, **Fig. 10c-d.** The main factor influencing the grains texture developed from recrystallisation is definitely the nucleation mechanism. The shear bands found within and across the columnar grains after rolling represent a strong conglomeration of crystals defects which could act as nucleation sites. Nucleation of strain-free grains at shear bands has been observed in many metals including copper and its alloys [44,45] and aluminium[46]. In the case analysed, the new recrystallised grains have a completely different orientation when compared to the deformed parent metal. The recrystallization orientation may correspond to the randomly orientated defects/nuclei placed within the slip bands.

*4.3 Porosity*

Scattered pores were distributed across the bulk of samples 1R and 2R (**Fig. 6a-c**) with some flattened oval pores found on the upper part of sample 2R (**Fig. 7a**). The presence of porosity was particularly reduced within the recrystallised region of sample 3R (**Fig. 6d**). This phenomenon has been already reported in aluminium structures deposited via WAAM and rolled with a load of 45 kN [47]. The pore closure was also reported by Toda et al. [48] and Chaijaruwanich et al. [49] respectively for the cold and hot rolling of aluminium structures. The gas entrapped within the pore was considered to be mainly dissociated atomic oxygen. When the load was applied, the pores slightly changed their shapes and the surrounding material was characterised by a higher dislocation density. This could have led to entrapment of atomic gas in dislocation and vacancies induced by the rolling step. Additionally, the elimination of the porosity could have been promoted by the dissolution of the gaseous atoms within the bulk of the new recrystallised grains during their growth.

# 5  Conclusions

The impact of cold vertical rolling with a load of 50 kN on the porosity, the microstructure and the texture of an AM tantalum component has been studied. A fundamental study has been performed in order to understand the change in microstructure and hardness that occurred after every step of the process. Three main samples corresponding to three sequential steps were used in this study: the as-deposited material (sample 1R), as-deposited + rolling (sample 2R), as-deposited + rolling + deposition (sample 3R). The sample 3R showed an improved microstructure already with only one rolling pass and one subsequent deposition. Sample 4R was used for the evaluation of the texture of the refined microstructure. The main findings of the study are summarised as follows:

- By combining the effect of cold rolling and the successive layer deposition, it is possible to achieve recrystallisation in AM tantalum components;

- The refined region found in sample 3R exactly corresponded to the peak in hardness of sample 2R. Thus, the evaluation of strain distribution using hardness map helped to understand the depth of the deposit affected the single rolling step;

- The elimination of pores due to rolling has been observed and explained by the presence of entrapped oxygen atomic gas within dislocation and vacancies, and/or by the dissolution of the oxygen within the bulk of the new recrystallised grains during their growth;

- The new strain-free grains presented an equiaxed microstructure and a complete random texture.

## Acknowledgement

The authors wish to acknowledge financial support from AWE and the valuable scientific contribution of Geoff Shrimpton (AWE) and Tim Rogers (AWE).